\documentclass[a4paper,12pt]{article}

\usepackage[hmargin=.7in,vmargin=1.1in]{geometry}
\usepackage{indentfirst}
\usepackage{amsfonts}
\usepackage{mathrsfs}
\usepackage{amsmath}
\usepackage{amssymb}
\usepackage{authblk}
\usepackage{cite}
\usepackage{xcolor}
\usepackage{mathtools}
\usepackage{tensor}
\usepackage{physics}
\usepackage{graphicx}
\usepackage{bm}
\usepackage{tikz}
\usepackage{upgreek}
\usepackage{braket}
\usepackage{color,soul}
\usepackage{csquotes}
\usepackage{caption}
\usepackage{subcaption}
\usepackage{tabularx}
\usepackage{simplewick}
\usepackage[section]{placeins}
\usepackage{enumerate}
\usepackage{comment}
\newcommand{\pp}{\uppi}

\newcommand\mi{\mathrm{i}}
\newcommand\me{\mathrm{e}}

\newcommand{\dif}{\mathrm{d}}

\usepackage[bookmarksnumbered=true,bookmarksopen=true]{hyperref}
 \hypersetup{colorlinks,
             linkcolor=[rgb]{0,0.3,0.6}, 
             citecolor=[rgb]{0,0.3,0.6}, 
             urlcolor=[rgb]{0,0.3,0.6}}

\title{\Large One-loop corrections to infrared GWs is forbidden by symmetries}

    \author[1,2]{Cheng-Jun Fang\thanks{fangchengjun@itp.ac.cn}}
    \author[1,2]{Han-Wen Hu\thanks{huhanwen@itp.ac.cn}}
    \author[1,2,3]{Zong-Kuan Guo\thanks{guozk@itp.ac.cn}}

\affil[1]{\normalsize{\em Institute of Theoretical Physics, Chinese Academy of Sciences, P.O. Box 2735, Beijing 100190, China}}	

\affil[2]{\normalsize{\em School of Physical Sciences, University of Chinese Academy of Sciences, No.19A Yuquan Road, Beijing 100049, China}}	

\affil[3]{\normalsize{\em School of Fundamental Physics and Mathematical Sciences, Hangzhou Institute for Advanced Study, University of Chinese Academy of Sciences, Hangzhou 310024, China}}

\numberwithin{equation}{section}

\date{}

\begin{document}

\maketitle

\begin{abstract}
Small-scale scalar perturbations amplified during inflation can induce primordial gravitational waves through tensor-scalar interactions. A long-standing controversial issue is whether the one-loop corrections to tensor perturbations exist on large scales. Firstly, we demonstrate through direct one-loop calculations that one-loop corrections cancel each other out on large scales. We then proceed from the symmetry of the interacting system and directly prove, based on the Ward identity, the absence of one-loop corrections on large scales—without the need for specific loop diagram calculations. This is consistent with the results we previously obtained for scalar perturbations.
\end{abstract}

\section{Introduction}
Introducing a period of inflation at the very beginning of the Universe \cite{Guth:1980zm,Starobinsky:1980te} is the most natural scenario to explain the behavior of primordial perturbations. Observations of the Cosmic Microwave Background (CMB) \cite{Planck:2018jri,Planck:2019kim,ACT:2025fju} and large-scale structures \cite{BOSS:2016wmc,DESI:2024mwx,DESI:2024hhd} indicate that primordial curvature perturbations are Gaussian perturbations with a nearly scale-invariant spectrum, which is precisely the prediction of the slow-roll inflation model \cite{Starobinsky:1982ee,Mukhanov:1990me}. In slow-roll inflation, the inflaton always evolves along the slow-roll attractor, however, this is not necessarily the end of the story. 

There are various indications that the slow-roll conditions may be violated on smaller scales. A series of observations related to primordial black holes \cite{Green:2020jor,Carr:2020xqk,KAGRA:2021duu,DeLuca:2020agl}and the stochastic gravitational wave background \cite{NANOGrav:2023hde,NANOGrav:2023gor,Zic:2023gta,Reardon:2023gzh,EPTA:2023fyk,EPTA:2023sfo,Xu:2023wog} all suggest that, compared to the CMB scale, the spectrum of primordial curvature perturbations on small scales is likely to have been amplified by several orders of magnitude. 
Therefore, a series of non-slow-roll inflation models have been proposed to generate such large primordial perturbations, such as ultra-slow-roll (USR) \cite{Namjoo:2012aa,Garcia-Bellido:2017mdw,Germani:2017bcs,Motohashi:2017kbs,Xu:2019bdp,Fu:2019vqc} or parametric resonance \cite{Cai:2018tuh,Cai:2019jah,Cai:2019bmk}.

Most studies on these models have been confined to the linear level, however, the inherent amplification feature of these models implies that they are likely to exhibit strong nonlinear effects, which can be explicitly shown in loop corrections and lattice simulations \cite{Inomata:2022yte,Caravano:2024tlp,Kristiano:2022maq}.
Among the issues related to nonlinear effect, the most noteworthy one is whether the amplification of perturbations on small scales will modify large-scale perturbations through interactions, thereby affecting our observational results in the CMB frequency band.

It is quite confusing that direct loop calculations suggest that both scalar \cite{Cheng:2021lif,Kristiano:2023scm,Kristiano:2022maq} and tensor perturbations \cite{Ota:2022hvh,Ota:2022xni} can receive relatively large one-loop corrections on super-horizon scales. This not only contradicts the intuition that physics should be local but also deviates from the conclusions of many theorems for similar situations \cite{Assassi:2012et,Senatore:2012ya,Tanaka:2015aza}. Many works have been dedicated to clarifying this issue \cite{Firouzjahi:2023aum,Firouzjahi:2023bkt,Fumagalli:2023zzl,kristianoComparingSharpSmooth2024,Choudhury:2023vuj,Franciolini:2023agm,Iacconi:2023ggt,Davies:2023hhn,Maity:2023qzw,Tasinato:2023ukp,Cheng:2023ikq,Ballesteros:2024zdp,Sheikhahmadi:2024peu,Riotto:2023hoz,Tada:2023rgp,Kawaguchi:2024lsw,Fang:2025tgk}, and this process has greatly deepened our understanding of inflationary field theory. Refs. \cite{Fumagalli:2024jzz,Kawaguchi:2024rsv} pointed out the connection between symmetry and the conservation of super-horizon perturbations, Refs. \cite{Firouzjahi:2025gja,Firouzjahi:2025ihn} took a first step towards n-loop calculations, while Refs.\cite{Inomata:2024lud,Fang:2025vhi,Inomata:2025bqw,Braglia:2025cee,Braglia:2025qrb} showed that properly accounting for backreaction effects in one-loop calculations naturally cancels out super-horizon corrections. Recently, our work \cite{Fang:2025kgf} further analyzed how backreaction preserves the system's symmetries and generalized the conclusion of one-loop conservation to all loop orders directly from the Ward identities. Beyond these, Ref.\cite{Kristiano:2025ajj} explored potential issues arising from renormalization in inflationary field theory.

In this paper, as a direct extension of the results on scalar perturbations, we aim to derive similar conclusions for tensor perturbations. It is also necessary‌ since large scale tensor perturbations directly correspond to primordial gravitational waves (GWs) which can give rise to detectable B-mode polarizations in CMB \cite{BICEP2:2014owc,BICEP:2021xfz,Seljak:1996gy}. We limit our research scope to non-attractor single-field inflation models, particularly a class of models represented by parametric resonance, which is also the most natural mechanism for generating observable primordial gravitational waves. In the first part, we provide a more explicit introduction to the scenarios under consideration.\footnote{This part of the content is also included in recent literature \cite{Ema:2025ftj}. We independently completed these calculations before the publication of Ref.\ \cite{Ema:2025ftj}, so we have also presented our calculation process here. Additionally, they mentioned that this conservation may be related to symmetry, but no rigorous proof was provided through the Ward identity which is added in our paper.} In the second part, we prove through direct calculations that one-loop perturbations cancel each other on large scales. In the third part, we present the symmetries satisfied by the theory and directly demonstrate the connection between such cancellation and symmetry starting from the Ward identity.
\section{Actions and linear order solutions}
In this paper, we consider the one-loop corrections to primordial GWs generated by single-field inflation with a potential $V(\phi)$ in the spatially flat gauge. The potential is chosen such that inflation satisfies the following scenario: an intermediate process between two slow-roll (SR) periods, namely SR\,-intermediate period-\,SR. We further assume that the first SR parameter $\epsilon$ remains small throughout the entire inflationary process, which is applicable to both ultra-slow-roll and parametric resonance scenarios~\cite{Inomata:2024lud,Inomata:2022yte,Cai:2019bmk}.
In this gauge, the metric can be written via the ADM decomposition as \cite{Maldacena:2002vr}
\begin{align}
 \dif s^2&=a^2\left[-N^2 \dif \tau^2+\me^{h_{i j}}\left(\dif x^i+N^i \dif t\right)\left(\dif x^j+N^j \dif t\right)\right] \nonumber \\
& \simeq a^2\left[-\dif \tau^2+\me^{h_{i j}} \dif x^i \dif x^j\right].
\end{align} 
We have neglected the lapse and shift since they are suppressed by $\epsilon$, so the action can be written as
\begin{align} \label{action}
 S & =S_{\rm GR}+\int \dif \tau \dif^3 x\ \sqrt{-g}\left[-\frac{1}{2} g^{\mu \nu} \partial_\mu \phi \partial_\nu \phi-V(\phi)\right]\nonumber \\
& =S_{\rm GR}+\int \dif \tau \dif^3 x\ a^4\left[\frac{1}{2} a^{-2} \phi^{\prime 2}-\frac{1}{2}\left(\me^{-h}\right)^{i j} a^{-2} \partial_i \phi \partial_j \phi-V(\phi)\right],
\end{align}
where the $S_{\rm GR}$ represents the Einstein-Hilbert action of gravity. Generally speaking, the effect of tensor self-interaction is much smaller than the GWs induced by field perturbations. Therefore, in the problem of tensor one-loop corrections, only the linear-order action $S_h$ in $ S_{\text{GR}}$ is important for us
\begin{equation}
S_h=\frac{M_{\rm p}^2}{8} \int \dif \tau \dif^3 x \ a^2\left[h_j^{i \prime} h_i^{j \prime}-\left(\partial_k h_j^i\right) \partial^k h_i^j\right].
\end{equation}
Expanding the inflaton field part of the action in Eq.(2.2) \ref{action} yields the free Lagrangian of field perturbations $S_{\delta \phi}$ and two interaction terms $S_{\rm int}$ related to tensor one-loop corrections
\begin{equation}
S_{\delta \phi}=\int \dif \tau \dif^3 x\ a^2\left(\frac{1}{2} \delta \phi^{\prime 2}-\frac{1}{2}\left(\partial_i \delta \phi\right)^2\right)-\frac{1}{2} V_{(2)} \delta \phi^2,
\end{equation}
\begin{equation}
S_{\rm int}=\int \dif \tau \dif^3 x \ a^2\left[\frac{1}{2} h^{i j}-\frac{1}{4} h^{i k} h_k^j\right] \partial_i \delta \phi  \partial_j \delta \phi\equiv S_3+S_4.
\end{equation}
With quadratic action $S_{h}$ and $S_{\delta\phi}$, we can first solve out the evolution of field operators in the interaction picture
\begin{equation}
\delta \phi=\int \frac{\dif^3 \mathbf{q}}{(2 \pp)^3} \me^{\mi \mathbf{q} \cdot \mathbf{x}}\left(u_q \hat{a}_{\mathbf{q}}+u_q^* \hat{a}_{-\mathbf{q}}^{\dagger}\right),
\end{equation}
\begin{equation}
h_{i j}=\int \frac{\dif^3 \mathbf{q}}{(2 \pp)^3} \me^{\mi \mathbf{q} \cdot \mathbf{x}} \sum_{s= \pm 2} e_{i j}^s(\hat{q})\left(v_q \hat{b}_{\mathbf{q}}^s+v_q^* \hat{b}_{-\mathbf{q}}^{s \dagger}\right),
\end{equation}
where $e_{i j}^s(\hat{q})$ is the polarization tensor while $u_q$ and $v_q$ are the mode functions of each k-mode. From the free field actions, we can directly obtain the equation of motion of these mode functions,
\begin{align}
\begin{aligned}
    \left[\frac{\dif^2}{\dif \tau^2}+2 \mathcal{H} \frac{\dif}{\dif \tau}+k^2+a^2 V_{(2)}\right] u_k&=0\equiv \hat{N}_k u_k,\\
    \left[\frac{\dif^2}{\dif \tau^2}+2 \mathcal{H} \frac{\dif}{\dif \tau}+k^2\right] v_k&=0.
\end{aligned}
\end{align}
We introduce two Green’s functions $G^h_q$ and $G_q$ for tensor and scalar respectively to 
simplify the expressions later in this paper
\begin{align}
		\left[h_{\mathbf{p}}^s\left(\tau^{\prime}\right), h_{\mathbf{q}}^{s_1}(\tau)\right]^\prime&=\left(v_q\left(\tau^{\prime}\right)v_q^*(\tau)-v_q^*\left(\tau^{\prime}\right) v_q(\tau)\right)\delta^{ss_1} \nonumber\\
		&\equiv \frac{4}{M_{\rm p}^2} \frac{\mi}{a(\tau^\prime)^2}G_q^h\left(\tau ; \tau^{\prime}\right)\delta^{ss_1},
	\end{align}
    \begin{align}
		\left[\delta\phi_{\mathbf{p}}\left(\tau^{\prime}\right), \delta\phi_{\mathbf{q}}(\tau)\right]^\prime&=\left(u_q\left(\tau^{\prime}\right)u_q^{*}(\tau)-u_q^{*}\left(\tau^{\prime}\right) u_q(\tau)\right) \nonumber\\
		&\equiv  \frac{\mi}{a(\tau^\prime)^2}G_q\left(\tau ; \tau^{\prime}\right).
	\end{align}
\section{Cancellation of the one-loop diagrams}
\begin{figure}[!htb]
    \centering
    \begin{subfigure}[b]{0.45\textwidth}
        \centering
        \begin{tikzpicture}[x=0.75pt,y=0.75pt,yscale=-1,xscale=1]
            \draw    (248.39,169.63) .. controls (250.04,167.96) and (251.71,167.95) .. (253.39,169.6) .. controls (255.06,171.26) and (256.73,171.25) .. (258.39,169.58) .. controls (260.05,167.91) and (261.72,167.9) .. (263.39,169.55) .. controls (265.06,171.2) and (266.73,171.19) .. (268.39,169.52) .. controls (270.05,167.85) and (271.72,167.84) .. (273.39,169.49) .. controls (275.06,171.14) and (276.73,171.13) .. (278.39,169.46) .. controls (280.05,167.79) and (281.72,167.78) .. (283.39,169.43) .. controls (285.06,171.08) and (286.73,171.07) .. (288.39,169.4) .. controls (290.05,167.73) and (291.72,167.72) .. (293.39,169.37) .. controls (295.06,171.02) and (296.73,171.01) .. (298.39,169.34) .. controls (300.05,167.67) and (301.72,167.66) .. (303.39,169.31) .. controls (305.06,170.97) and (306.73,170.96) .. (308.39,169.29) .. controls (310.05,167.62) and (311.72,167.61) .. (313.39,169.26) .. controls (315.06,170.91) and (316.73,170.9) .. (318.39,169.23) .. controls (320.05,167.56) and (321.72,167.55) .. (323.39,169.2) .. controls (325.06,170.85) and (326.73,170.84) .. (328.39,169.17) .. controls (330.05,167.5) and (331.72,167.49) .. (333.39,169.14) .. controls (335.06,170.79) and (336.73,170.78) .. (338.39,169.11) .. controls (340.05,167.44) and (341.72,167.43) .. (343.39,169.08) .. controls (345.06,170.73) and (346.73,170.72) .. (348.39,169.05) -- (350.18,169.04) -- (350.18,169.04) ;
            \draw [shift={(299.29,169.34)}, rotate = 359.67] [color={rgb, 255:red, 0; green, 0; blue, 0 }  ][fill={rgb, 255:red, 0; green, 0; blue, 0 }  ][line width=0.75]      (0, 0) circle [x radius= 2.01, y radius= 2.01]   ;
            \draw   (276.44,146.32) .. controls (276.44,133.59) and (286.76,123.27) .. (299.49,123.27) .. controls (312.22,123.27) and (322.54,133.59) .. (322.54,146.32) .. controls (322.54,159.05) and (312.22,169.37) .. (299.49,169.37) .. controls (286.76,169.37) and (276.44,159.05) .. (276.44,146.32) -- cycle ;
        \end{tikzpicture}
        \caption{corresponding to $\left\langle h_{-\mathbf{q}}^{r,(1)} h_{\mathbf{q}}^{s,(3b)}\right\rangle^{\prime}$}
        \label{fey2}
    \end{subfigure}
    \hfill 
    \begin{subfigure}[b]{0.45\textwidth}
        \centering
        \begin{tikzpicture}[x=0.75pt,y=0.75pt,yscale=-1,xscale=1]
            \draw    (227.99,149.57) .. controls (229.64,147.88) and (231.3,147.86) .. (232.99,149.51) .. controls (234.68,151.15) and (236.34,151.13) .. (237.99,149.44) .. controls (239.64,147.75) and (241.3,147.73) .. (242.99,149.37) .. controls (244.68,151.02) and (246.34,151) .. (247.99,149.31) .. controls (249.64,147.62) and (251.3,147.6) .. (252.99,149.24) .. controls (254.68,150.89) and (256.34,150.87) .. (257.99,149.18) .. controls (259.64,147.49) and (261.3,147.47) .. (262.99,149.11) .. controls (264.68,150.75) and (266.34,150.73) .. (267.99,149.04) .. controls (269.64,147.35) and (271.3,147.33) .. (272.99,148.98) -- (276.08,148.94) -- (276.08,148.94) ;
            \draw [shift={(276.08,148.94)}, rotate = 359.25] [color={rgb, 255:red, 0; green, 0; blue, 0 }  ][fill={rgb, 255:red, 0; green, 0; blue, 0 }  ][line width=0.75]      (0, 0) circle [x radius= 2.01, y radius= 2.01]   ;
            \draw    (326.08,148.94) .. controls (327.71,147.25) and (329.38,147.22) .. (331.07,148.86) .. controls (332.76,150.5) and (334.43,150.47) .. (336.07,148.78) .. controls (337.71,147.09) and (339.38,147.06) .. (341.07,148.7) .. controls (342.76,150.34) and (344.43,150.31) .. (346.07,148.62) .. controls (347.71,146.93) and (349.38,146.9) .. (351.07,148.54) .. controls (352.76,150.18) and (354.43,150.15) .. (356.07,148.46) .. controls (357.71,146.77) and (359.38,146.74) .. (361.07,148.37) .. controls (362.76,150.01) and (364.43,149.98) .. (366.07,148.29) -- (366.95,148.28) -- (366.95,148.28) ;
            \draw [shift={(326.08,148.94)}, rotate = 359.08] [color={rgb, 255:red, 0; green, 0; blue, 0 }  ][fill={rgb, 255:red, 0; green, 0; blue, 0 }  ][line width=0.75]      (0, 0) circle [x radius= 2.01, y radius= 2.01]   ;
            \draw   (276.08,148.94) .. controls (276.08,135.13) and (287.27,123.94) .. (301.08,123.94) .. controls (314.88,123.94) and (326.08,135.13) .. (326.08,148.94) .. controls (326.08,162.75) and (314.88,173.94) .. (301.08,173.94) .. controls (287.27,173.94) and (276.08,162.75) .. (276.08,148.94) -- cycle ;
        \end{tikzpicture}
        \caption{corresponding to $\left\langle h_{-\mathbf{q}}^{r,(1)} h_{\mathbf{q}}^{s,(3a)}\right\rangle^{\prime}$}
        \label{fey1}
    \end{subfigure}
    \caption{one-loop order tensor diagrams.}
    \label{fig:enter-label}
\end{figure}
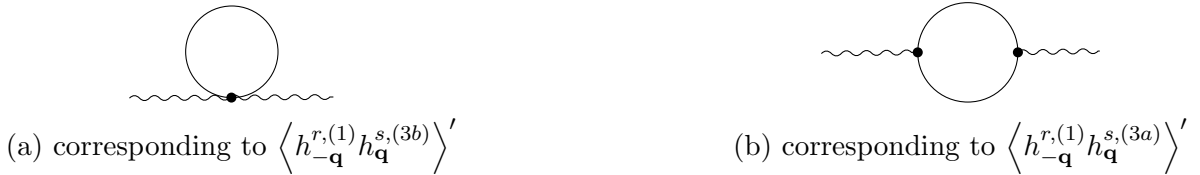

Our goal is to calculate the power spectrum of tensor perturbations
\begin{align}
     \left\langle \Omega\left|\sum_s h_{\mathbf{q},\rm H}^s(\tau) h_{\mathbf{q}^\prime,\rm H}^s(\tau)\right| \Omega\right\rangle\equiv\frac{(2 \pp)^3}{q^3} \delta(\mathbf{q}+{\mathbf{q}^\prime}) P_h,
\end{align}
where $h_{\mathbf{q},\rm H}^s(\tau)$ is an operator in the Heisenberg picture. A Heisenberg picture operator can always be expanded by free operators through the Dyson series
    \begin{align}
		h_{\mathbf{q},\rm H}^s(\tau)&=F\left(\tau ; \tau_i\right)^{\dagger} h_{\mathbf{q}}^s F\left(\tau ; \tau_i\right) \nonumber\\
		&=h_{\mathbf{q}}^s(\tau)+\mi \int_{\tau_i}^\tau \dif \tau^{\prime}\left[H_{\mathrm{int}, {\rm I}}\left(\tau^{\prime}\right), h_{\mathbf{q}}^s(\tau)\right]-\int_{\tau_i}^\tau \dif \tau^{\prime} \int_{\tau_i}^{\tau^{\prime}} \dif \tau^{\prime \prime}\left[H_{\mathrm{int}, {\rm I}}\left(\tau^{\prime \prime}\right),\left[H_{\mathrm{int}, {\rm I}}\left(\tau^{\prime}\right), h_{\mathbf{q}}^s\right]\right] \nonumber\\
         & +\cdots.
	\end{align}
With this formula we expand the Heisenberg picture operators $h_{\mathbf{q},\rm H}^s(\tau)$ into a polynomial consisting of interaction picture operators. Then, we categorize the terms according to the power of the operators in this polynomial
\begin{equation}
		h_{\mathbf{q},\rm H}^s(\tau)\equiv h_\mathbf{q}^{s,(1)}+h_\mathbf{q}^{s,(2)}+h_\mathbf{q}^{s,(3)}+\cdots.
\end{equation}
The category here is done with respect to the number of interaction picture operators in each term, for example, $\mi \int_{\tau_i}^\tau \dif \tau^{\prime}\left[H^{(4)}_{ {\rm I}}\left(\tau^{\prime}\right), h_{\mathbf{q}}^s(\tau)\right]$ includes terms like \(h\delta\phi^2\), which classifies it as a third-order term. As mentioned in \cite{Ota:2022hvh,Ota:2022xni,Gong:2019yyz,Fang:2025tgk}, the power spectrum of the so-called induced GWs $\left\langle h_{\mathbf{k}}^{(2)} h_{\mathbf{-k}}^{(2)}\right\rangle$ is suppressed by $k^3$ in the IR limit, and the dominant IR contribution is given by $\left\langle h_{\mathbf{k}}^{(1)} h_{\mathbf{-k}}^{(3)}\right\rangle$. This property leads us to calculate $h_{\mathbf{q}}^{(3)}$ in detail, and the results are given by
 \begin{subequations}
	\begin{align}
		h_\mathbf{q}^{s,(3)}=&h_\mathbf{q}^{s,(3a)}+h_\mathbf{q}^{s,(3b)},&\\
        h_\mathbf{q}^{s,(3a)} =&\frac{2}{M_{\rm p}^2}\int_{\tau_i}^\tau \dif \tau^{\prime}\int_{\tau_i}^{\tau^\prime}\dif \tau^{\prime\prime} G_q^G\left(\tau ; \tau^{\prime}\right) \nonumber\\
		&\times   \int \frac{\dif^3 \mathbf{p} \dif^3 \mathbf{k}}{(2 \pp)^6}e_{i j}^{s*}(\hat{q})\mathbf{p}_i \mathbf{p}_j  G_p\left(\tau^\prime ; \tau^{\prime\prime}\right) \sum_{s_1} h_\mathbf{k}^{s_1}(\tau^{\prime\prime}) \nonumber\\
		&\times e_{mn}^{s_1}(\hat{k})\mathbf{p}_m \mathbf{p}_n  \{\delta\phi_{\mathbf{p}-\mathbf{k}}(\tau^{\prime\prime}) \delta\phi_{\mathbf{q}-\mathbf{p}}(\tau^{\prime})\}, \\
        h_\mathbf{q}^{s,(3b)}=&-\frac{2}{M_{\rm p}^2}\int_{\tau_i}^\tau \dif \tau^{\prime}G_q^G\left(\tau ; \tau^{\prime}\right)\int \frac{\dif^3 p \dif^3 k}{(2 \pp)^6} \sum_{s_2}e_{i k}^{s*}(\hat{q}) \nonumber\\
		&\times e_{k j}^{s_2}(\hat{p}) \mathbf{k}_i(\mathbf{q}-\mathbf{k})_j \delta\phi_\mathbf{k}(\tau^{\prime})\delta\phi_{\mathbf{q}-\mathbf{p}-\mathbf{k}}(\tau^{\prime})h_\mathbf{p}^{s_2}(\tau^{\prime}),
	\end{align}
\end{subequations}
from which we can arrive at the correlation functions through Wick contractions. The results of those two terms are listed as follows,
\begin{align}
 \left\langle h_{-\mathbf{q}}^{r,(1)} h_{\mathbf{q}}^{s, (3b)}\right\rangle ^{\prime}& =-\frac{2}{M_{\rm p}^2} \int^\tau \dif \tau^{\prime} G_q^h\left(\tau; \tau^{\prime}\right) \int \frac{\dif^3 \mathbf{k}}{(2 \pp)^3} e_{i k}^{s *}(\hat{q}) e_{k j}^r(\hat{q}) \mathbf{k}_i(\mathbf{q}-\mathbf{k})_j u_{k}\left(\tau^{\prime}\right) u_{k}^*\left(\tau^{\prime}\right) v_q(\tau) v_q^*\left(\tau^{\prime}\right)\nonumber\\
 &=-\frac{2}{3 M_{\rm p}^2} \int^\tau \dif \tau^{\prime} G_q^h\left(\tau; \tau^{\prime}\right) \delta_{i j}\left[\int \frac{\dif^3 \mathbf{k}}{(2 \pp)^3} k^2\left|u_k\left(\tau^{\prime}\right)\right|^2\right]e_{i k}^{s *}(\hat{q}) e_{k j}^r(\hat{q}) v_q(\tau) v_q^*\left(\tau^{\prime}\right)\nonumber\\
 &=-\frac{2}{3 M_{\rm p}^2} \delta^{r s} \int^\tau \dif \tau^{\prime} G_q^h\left(\tau; \tau^{\prime}\right) v_q(\tau) v_q^*\left(\tau^{\prime}\right)\left[\int \frac{\dif^3 \mathbf{k}}{(2 \pp)^3} k^2\left|u_k\left(\tau^{\prime}\right)\right|^2\right],
\end{align}
\begin{align}
    \left\langle h_{-\mathbf{q}}^{r,(1)} h_{\mathbf{q}}^{s,(3a)}\right\rangle^{\prime}&=\frac{2}{M_{\rm p}^2} \int^\tau \dif \tau^{\prime} \int^{\tau^{\prime}} \dif \tau^{\prime \prime} G_q^h\left(\tau; \tau^{\prime}\right) \int \frac{\dif^3 \mathbf{p}}{(2 \pp)^3} e_{i j}^{s *}\mathbf{p}_i \mathbf{p}_j G_p\left(\tau^{\prime}; \tau^{\prime \prime}\right) v_q(\tau) v_q^*\left(\tau^{\prime \prime}\right) e_{m n}^r(\hat{q}) \mathbf{p}_m \mathbf{p}_n \nonumber \\
    & \quad \times 2 \Re\left[u_{q-p}\left(\tau^{\prime}\right) u_{q-p}^*\left(\tau^{\prime \prime}\right)\right]\nonumber \\
    &=\frac{2}{M_{\rm p}^2} \int^\tau \dif \tau^{\prime} G_q^h\left(\tau; \tau^{\prime}\right) \delta^{r s} \int \frac{\dif^3 \mathbf{p}}{(2 \pp)^3} \left| e_{i j}^{s *}(\hat{q}) \mathbf{p}_i \mathbf{p}_j\right|^2 \int ^{\tau^{\prime}}\dif \tau^{\prime \prime} G_p\left(\tau^{\prime}; \tau^{\prime \prime}\right) v_q(\tau) v_q^*\left(\tau^{\prime \prime}\right) \nonumber \\
    &\quad \times 2 \Re\left[u_{q-p}\left(\tau^{\prime}\right) u_{q-p}^*\left(\tau^{\prime \prime}\right)\right].
\end{align}
These two terms correspond to \ref{fey2} and \ref{fey1} respectively.
The results of these two diagrams do not seem to be related. However, as long as the scalar mode function satisfies the equation of motion $\hat{N}_k u_k=0$, it will satisfy an integral expression
\begin{equation} \label{integral}
\frac{\dif}{\dif \log k}\left|u_k\right|^2=-2 k^2 \int^{\tau_1} \dif \tau G_k\left(\tau_1; \tau\right) 2 \Re\left[u_k(\tau) u_k^*\left(\tau_1\right)\right].
\end{equation}
The specific proof will be provided in Appendix \ref{CR}. With the help of this integral relation, we can further simplify the results in Fig.\ref{fey1} as
\begin{align}
    \left\langle h_{-\mathbf{q}}^{r,(1)} h_{\mathbf{q}}^{s,(3a)}\right \rangle ^{\prime}&=\frac{2}{M_{\rm p}^2} \int^\tau \dif \tau^{\prime} G_q^r\left(\tau; \tau^{\prime}\right) \delta^{r s} \int \frac{\dif^3 \mathbf{p}}{(2 \pp)^3}\left| e_{i j}^{s *}(\hat{q}) \mathbf{p}_i \mathbf{p}_j\right|^2 v_q(\tau) v_q^*(\tau) \left(-\frac{1}{2 p^2} \frac{\dif}{\dif \log p}\left|u_p\right|^2\right)\nonumber \\
    &=-\frac{\delta^{r s}}{M_{\rm p}^2} \int^\tau \dif \tau^{\prime}\  G_q^s\left(\tau; \tau^{\prime}\right) \frac{2}{15}\left|v_q(\tau)\right|^2 \int \frac{\dif^3 \mathbf{p}}{(2 \pp)^3} p^2 \frac{\dif}{\dif \log p}\left|u_p\right|^2\nonumber\\
    &= \frac{2 \delta^{r s}}{3 M_{\rm p}^2} \int^\tau \dif \tau^{\prime}\  G_q^s\left(\tau; \tau^{\prime}\right) \left|v_q(\tau)\right|^2 \int \frac{\dif^3 \mathbf{p}}{(2 \pp)^3} p^2 \left|u_p\right|^2.
\end{align}
In this process, we assume by default that when the interaction occurs, the mode corresponding to the external momentum $q$ is already super-horizon, so its mode function has also been frozen.
We immediately notice that after some simple algebra, these two expressions coincide
\begin{equation}
\left\langle h_{-\mathbf{q}}^{r,(1)} h_{\mathbf{q}}^{s,(3a)}\right\rangle^{\prime}=-\left\langle h_{-\mathbf{q}}^{r,(1)} h_{\mathbf{q}}^{s,(3b)}\right\rangle^{\prime},
\end{equation}
which means that super-horizon corrections totally cancels each other at one-loop order. 
\section{Symmetry and Ward Identities}
In the previous chapters, we have proven through direct calculations that the contributions of tensor one-loop diagrams on large scales cancel each other out. However, Ref \cite{Fang:2025tgk} points out that for scalar perturbations, such cancellation between loop diagrams is directly caused by the Ward identity derived from symmetry. Therefore, we hope to use a similar approach to establish the relationship between the super-horizon conservation of tensor perturbations and symmetry. 
To achieve this, we start our analysis from the symmetry of the action. The following four terms of the action are important at the one-loop order,
\begin{equation}
S=S_h+S_{\delta \phi}+S_3+S_4.
\end{equation}
We notice that if we consider the following transformation \cite{Hinterbichler:2013dpa,Hinterbichler:2012nm}, that is, adding a constant tensor to $ h_{ij}$  while performing a coordinate transformation,
\begin{align}
\begin{aligned}
    \tilde{x}^i&=x^i-M_{\;\; j}^i x^j,\\
    \tilde{h}_{i j}(\tilde{x})&=h_{i j}(x)+2 M_{i j},\\
    \delta\tilde{\phi}(\tilde{x})&=\delta\phi(x),
\end{aligned}
\end{align}
where the tensor $M_{ij}$ fulfills $M_{i j}=M_{j i}$ and $M_{i i}=0$, then the action remains unchanged after that transformation,
\begin{align}  
S & =S_{\delta \phi}+\int \dif \tau \dif^3 xa^2  M^{i j} \partial_i \delta \phi \partial_j \delta \phi\nonumber \\ 
&\quad +S_h+S_3-\int \dif \tau \dif^3 x\ \left[M^{i j} a^2 \partial_i \delta \phi \partial_j \delta \phi + a^2h^{i j} M^k_{\;\; j} \partial_i \delta \phi \partial_k \delta \phi\right]\nonumber\\
&\quad +S_4+\int \dif \tau \dif^3 x\ a^2 M^{i k} h_k{ }^j \partial_i \delta \phi \partial_j \delta \phi +...\nonumber\\
&\simeq S_{\delta \phi}+S_h+S_3+S_4.
\end{align}
Each term's variation generated after the transformation cancels each other out, ultimately keeping the total action invariant. The operator form of the Ward identity associated with this symmetry can be obtained through integrating the correlation function form of Ward identities~\cite{Assassi:2012zq,Assassi:2012et},
\begin{equation}\label{WI}
\mi\left[\hat{Q}, \hat{h}_{i j}\right]=-\delta \hat{h}_{i j}.
\end{equation}
where $\delta \hat{h}_{i j}$ is defined as
\begin{equation}
\delta \hat{h}_{i j}=M_{\;\; l}^k x^l \partial_k \hat{h}_{i j}+2 M_{i j}.
\end{equation}
To extract the constraints on the two-point correlation function, we take the expectation value of both sides of Eq.\ \eqref{WI}
\begin{equation}
\langle\Omega \mid -\delta h_{i j} \mid \Omega\rangle=i\langle\Omega|\left[\hat{Q}, \hat{h}_{i j}\right]|\Omega\rangle.
\end{equation}
The average of the left side can be obtained directly, while the right side requires further evaluation. We first consider the simultaneous eigenstates of the field configuration $|\phi\rangle \equiv\left|h_{i j}, \delta \phi\right\rangle$ which is defined as 
\begin{align}
\begin{aligned}
\hat{h}_{i j}(x)|\phi\rangle=h_{i j}(x)|\phi\rangle,\\
\delta\hat{\phi}(x)|\phi\rangle=\delta\phi(x)|\phi\rangle.
\end{aligned}
\end{align}
Since they form a complete set of bases, we can insert them into the right side as a unit operator
\begin{equation} \label{EV}
-2 M_{i j}=\mi \int \mathcal{D} h_{k l} \mathcal{D} \delta \phi\ \left\langle\Omega\right| \hat{Q}\left|\phi\right\rangle\left\langle\phi\right| \hat{h}_{i j} \left| \Omega\right\rangle-\text{c.c.}.
\end{equation}
Therefore, this problem is transformed into finding the matrix element $\left\langle\Omega\right| \hat{Q}\left|\phi\left(\tau_i\right)\right\rangle$, which leads us to focus on the transformation property of $|\phi\rangle$. Verifying the following relation, 
\begin{align}
 \left(\hat{h}_{i j}+\delta \hat{h}_{i j}\right)(1-\mi \hat{Q})|\phi\rangle & =h_{i j}|\phi\rangle+\delta \hat{h}_{i j} |\phi \rangle+\mi\left[\hat{Q}, \hat{h}_{i j}\right] |\phi\rangle-\mi \hat{Q} h_{i j}|\phi\rangle \nonumber \\
& =h_{i j}(1-\mi \hat{Q})|\phi\rangle \equiv h_{i j}|\psi\rangle,
\end{align}
we notice that $|\psi\rangle$ are also field configuration eigen states,
\begin{align}
    \left[\hat{h}_{i j}\left(x^k+M_{\;\; l}^k x^l\right)+2 M_{i j}\right] | \psi\rangle &=h_{i j}\left(x^k\right)|\psi\rangle \nonumber\\
    \hat{h}_{i j}\left(x^k\right)|\psi\rangle&=\left[h_{i j}\left(x^k-M^k_{\;\; l}x^k\right)-2 M_{i j}\right]|\psi\rangle\equiv \tilde{h}_{i j}|\psi\rangle.
\end{align}
As a result, the matrix element $\left\langle\Omega\right| \hat{Q}\left|\phi\right\rangle$ actually corresponds to the variation of the wave function $\left\langle\Omega \mid \phi\right\rangle$. However, the cosmic wave function generally contains complex non-Gaussianity terms unless these eigen-states are taken at early times when the interactions can be neglected. The early time wave functions can be written as
\begin{equation}
\left\langle\Omega \mid \phi_i\right\rangle \propto \exp \left[-\int \frac{\dif^3 \mathbf{k}}{(2 \pp)^3} \frac{1}{4} D_k h_{i j}(\mathbf{k}) h^{i j}(-\mathbf{k})\right],
\end{equation}
herein, we have omitted the wave function of the scalar part because it does not contribute to the final result, and the relevant proof can be found in the appendix \ref{V}. From the structure of the wave function, we can see that it is important to calculate the transformation of the Fourier coefficient
\begin{align}
\tilde{h}_{i j} & =-2 M_{i j}+\int \frac{\dif^3 \mathbf{p}}{(2 \pp)^3} \me^{i p_k\left(x^k-M_{\;\; l}^k x^l\right)} h_{i j}(\mathbf{p})\nonumber \\
& =-2 M_{i j}+\int \frac{\dif^3 \tilde{\mathbf{p}}}{(2 \pp)^3} \me^{i \tilde{p}_k x^k} h_{i j}\left(\tilde{p}_k+\tilde{p}_l M_{\;\; k}^l\right),
\end{align}
where $\tilde{p}_k$ is defined as $p_k-p_l M_{\;\; k}^l$, from which we can infer that 
\begin{equation}
\tilde{h}_{i j}\left(p_k\right)=h_{i j}\left(p_k+p_l M_{\;\; k}^l\right)-2 M_{i j} \delta(\mathbf{p}).
\end{equation}
The transformed wave function can thus be written as
\begin{align}
\left \langle \Omega \mid \psi \right \rangle &\propto \exp \left[-\int \frac{\dif^3 \mathbf{k}}{(2 \pp)^3} \frac{1}{4} \tilde{h}_{i j}(\mathbf{k}) D_k \tilde{h}_{i j}(-\mathbf{k})\right] \nonumber\\
&=\exp \left[-\int \frac{\dif^3 \tilde{\mathbf{k}}}{(2 \pp)^3} \frac{1}{4} h_{i j}(\tilde{\mathbf{k}}) D_{\tilde{k}} h_{i j}(-\tilde{\mathbf{k}})\right](1+M_{i j} D_0 h_{i j}(0)+V ),
\end{align}
where we have applied a variable substitution $\tilde{k}_k=k_k+k_l M_{\;\; k}^{l}$ and defined $D_k=D_{\tilde{k}}+\Delta D_k$. Meanwhile, the additional term arising from $\Delta D_k$ have been incorporated into $V$, and it's specific expression is as follows
\begin{equation}
V=-\int \frac{\dif^3 \mathbf{k}}{(2 \pp)^3} \frac{1}{4} h_{i j}(\mathbf{k}) \Delta D_k h_{i j}(-\mathbf{k})
\end{equation}
We will show in the appendix \ref{V} that $V$ doesn't contribute to the final result, thus the transformed wave functions can be written as
\begin{equation}
\left \langle \Omega \mid \psi \right \rangle =\left(1+M_{i j} D_0 h_{i j}(0)\right)\left\langle\Omega \mid \phi_i\right\rangle.
\end{equation}
We noticed that the matrix element mentioned above is exactly the variation of the wave function
\begin{align}
\left\langle\Omega \mid i Q \mid \phi_i\right\rangle =-M_{i j} D_0 h_{i j}(0)\left\langle\Omega \mid \phi_i\right\rangle.
\end{align}
Substituting it in to Eq.\ \eqref{EV} and applying the Spectral decomposition of $\hat{h}_{k l}\left(0\right)\big|_{\tau_i}$ gives following expression
\begin{align}
   2 M_{i j}&=M_{k l} D_0\left\langle\Omega\right|\hat{h}_{k l}\left(0\right)\big|_{\tau_i} \hat{h}_{i j}(0)\left| \Omega\right\rangle+\text{c.c.}.
\end{align}
 To compare this result with our one loop calculations, we still need mode expansions
\begin{equation}
h_{i j}(0)=\int \dif \theta \dif \varphi \sin \theta \sum_s e_{i_j}^s(\hat{q}) h_{\hat{q}}^s.
\end{equation}
Since $M$ is an arbitrary tensor satisfying the symmetric and traceless conditions, we can prove that the equation holds for any mode, namely
\begin{equation}
2\left\langle h_{\hat{q}}\left(\tau_i\right) h_{-\hat{q}}\left(\tau_i\right)\right\rangle=\left\langle h_{\hat{q}}\left(\tau_i\right) h_{-\hat{q}}(\tau)\right\rangle+\text{c.c.},
\end{equation}
which exactly coincides with our one loop result. 
\section{Discussion}
In this paper, we focus on the behavior of one-loop GWs induced by inflaton perturbations in the super-horizon limit within non-attractor single-field inflation models. Through calculations, we find that small-scale sources do not generate infrared corrections to the tensor perturbation spectrum, and we directly derive this conservation law via the Ward identity based on the symmetry of this theory. This result is a direct extension of the theorem we obtained in our previous study on scalar perturbations, demonstrating the universality of such symmetry-protected infrared conservation.

It is crucial to emphasize the reason why previous loop calculations yielded large loop corrections on large scales. Our calculations reveal that the satisfaction of the equation of motion by the scalar perturbation mode function plays a crucial role throughout the computation. The parameterization of scalar mode functions in earlier works did not correspond to solutions of the equation of motion \cite{Ota:2022hvh,Ota:2022xni}, which can be clearly seen by calculating the Wronskian determinant. This result reminds us that mode functions cannot be arbitrarily parameterized.

Meanwhile, there are still many issues to be addressed. First, it should be noted that for scalar perturbations, we can obtain results at arbitrary loop orders, whereas in this paper, we only discuss up to the one-loop level. This is because tensor perturbations exhibit strong gauge dependence at nonlinear orders and we are not clear which gauge invariant quantity corresponds to the observable. I believe it is possible to construct a super-horizon conserved quantity for tensor perturbations, which only requires that the action satisfies the symmetry non-perturbatively under such a parameterization. However, we cannot well explain the physical significance of this construction, so we have not simply generalized this proof to N-loop orders. We will try to clarify the gauge problem in future works. Secondly, in the one-loop calculations of both scalar and tensor perturbations, we have found that the tree-level consistency relations play an important role in loop cancellations. As we know, consistency relations essentially stem from symmetries. Therefore, it is also highly meaningful to clarify the specific details of how consistency relations connect symmetries and loop cancellations, and we will continue to work on this.

\appendix
\section{Proof of the integral relation}\label{CR}
This appendix focus on the proof of Eq.\ \eqref{integral}, we start from the definition of a new function,
\begin{equation}
f_k\left(\tau_1, \tau_2\right) \equiv u_k\left(\tau_1\right) u_k^*\left(\tau_2\right)+u_k\left(\tau_2\right) u_k^*\left(\tau_1\right).
\end{equation}
Since $\hat{N}_k u_k=0$, we also have
\begin{equation}
\hat{N}_k\left(\tau_1\right) f_k=\hat{N}_k\left(\tau_2\right) f_k=0.
\end{equation}
From which we can construct two new functions
\begin{equation}
C_k\left(\tau_1, \tau_2\right) \equiv 2 k^2 \int^{\tau_1} \dif \tau\ G_k\left(\tau_1, \tau\right) f_k\left(\tau, \tau_2\right),
\end{equation}
\begin{equation}
g_k\left(\tau_1, \tau_2\right)\equiv\frac{\dif}{\dif \log k} f_k+C_k\left(\tau_1, \tau_2\right)+C_k\left(\tau_2, \tau_1\right).
\end{equation}
Recalling the basic formula of the Green's function method, we know that
\begin{equation}
\hat{N}_k\left(\tau_1\right) C_k\left(\tau_1, \tau_2\right)=2 k^2 f_k\left(\tau_1, \tau_2\right).
\end{equation}
It is also obvious that
\begin{equation}
\hat{N}_k\left(\tau_2\right) C_k\left(\tau_1, \tau_2\right)=2 k^2 \int^\tau \dif \tau\ G_k\left(\tau_1, \tau\right)\left[\hat{N}_k\left(\tau_2\right) f_k\right]=0.
\end{equation}
Taking the derivative of $\hat{N}_k u_k=0$ with respect to wave number gives
\begin{equation}
\hat{N}_k\left(\tau_1\right) \frac{\dif f_k}{\dif \log k}=-2 k^2 f_k=\hat{N}_k\left(\tau_2\right) \frac{\dif f_k}{\dif \log k}
\end{equation}
Combining all results above, we arrived at 
\begin{equation}
\left[\hat{N}_k\left(\tau_1\right)+\hat{N}_k\left(\tau_2\right)\right] g_k=0,
\end{equation}
which is a partial derivative equation. During the early era of the inflation, the mode functions can be approximated by
\begin{equation}
u_k=\mi \frac{H}{\sqrt{2 k^3}}(1+\mi k \tau) \me^{-\mi k \tau}.
\end{equation}
From this, we can explicitly verify that $g_k=0$ during the early epoch, where we have properly chosen the integration path. According to the uniqueness of solutions to partial differential equations, $ g_k = 0 $ is the only solution that satisfies the initial conditions, and thus it will hold throughout the entire process.

\section{Transformation of the wave function}\label{V}
In the calculations in the main text, we have consistently omitted the part of the wave function related to scalar perturbations, and at the same time, we have also discarded a term $V$. In this appendix, we will supplement these missing details. First, we note that under the symmetry transformation we are considering, not only do tensor perturbations change, but scalar perturbations also undergo corresponding transformations. Therefore, the Ward identity has an additional part
\begin{equation}
\mi[\hat{Q}, \delta \hat{\phi}]=-\delta \delta \hat{\phi}, \quad \delta \delta \hat{\phi}=M^k_{\;\; l} x^l \partial_k \delta \hat{\phi}.
\end{equation}
We also consider the transformation of $|\phi\rangle$
\begin{align}
    (\delta \hat{\phi}+\delta \delta \hat{\phi})(1-\mi \hat{Q})|\phi\rangle=\delta \phi|\phi\rangle+\mi[\hat{Q}, \delta \hat{\phi}]|\phi\rangle -\mi \hat{Q} \delta \phi|\phi\rangle+\delta \delta \hat{\phi}|\phi\rangle=(1-\mi \hat{Q}) \delta \phi|\phi\rangle,
\end{align}
thus $|\psi\rangle$ are also eigen-states of $\delta \hat{\phi}$
\begin{equation}
\delta \hat{\phi}\left(x^k\right)|\psi\rangle=\delta \phi\left(x^k-M^k_{\;\; l}x^l\right)|\psi\rangle=\delta \phi\left(\tilde{x}^k\right) \left| \psi\right\rangle.
\end{equation}
The total wave function contains a tensor part as well as a scalar part
\begin{equation}
\left\langle\Omega \mid \phi_i\right\rangle \propto\left\langle\Omega \mid h_{i j}\right\rangle_i\langle\Omega \mid \delta \phi\rangle_i,
\end{equation}
which is also of Gaussian form during early times  
\begin{equation}
\langle\Omega \mid \delta \phi\rangle_i \propto \exp \int \frac{\dif^3 \mathbf{k}}{(2 \pp)^3} \frac{1}{2} \varepsilon_k \delta \phi_{\mathbf{k}} \delta \phi_{-\mathbf{k}}.
\end{equation}
Similar to the tensor case, we also consider the behavior of the scalar part of the wave function under this transformation, and we find that
\begin{equation}
\left\langle\Omega \mid \psi\right\rangle=\langle\Omega \mid \delta \phi\rangle\left(1-\int \frac{\dif^3 \tilde{\mathbf{k}}}{(2 \pp)^3} \frac{1}{2} \Delta \varepsilon_k \delta \phi_{\tilde{\mathbf{k}}} \delta \phi_{-\tilde{\mathbf{k}}}\right),
\end{equation}
where we have defined $\varepsilon_k=\varepsilon_{\tilde{k}}+\Delta \varepsilon_k$, which is the variation of $\varepsilon_k$ after changing the integration variable from $k$ to $\tilde{k}_k=k_k+k_l M_{\;\; k}^{l}$, and it's explicit form can be written as $\Delta \varepsilon_k=-\frac{\dif \varepsilon_k}{\dif k} k_i k_j M^{i j}$. Substituting this expression into the r.h.s. of the Ward identity we have
\begin{align}
&\mi \int \mathcal{D} h_{k l} \mathcal{D} \delta \phi\ \langle\Omega| \hat{Q}\left|\phi_i\right\rangle\left\langle\phi_i\right| \hat{h}_{i j}|\Omega\rangle \nonumber\\
&=-\mi \int \mathcal{D} h_k \mathcal{D} \delta \phi \int \frac{\dif^3 \mathbf{k}}{(2 \pp)^3} \frac{1}{2} \Delta \varepsilon_k\langle\Omega| \delta \phi_{\mathbf{k}} \delta \phi_{-\mathbf{k}}\left|\phi_i\right\rangle\left\langle\phi_i\left|\hat{h}_{i_j}\right| \Omega\right\rangle\nonumber\\
&=-\mi \int \frac{\dif^3 \mathbf{k}}{(2 \pp)^3} \frac{1}{2} \Delta \varepsilon_k\langle\Omega| \delta \hat{\phi}_{\mathbf{k}} \delta \hat{\phi}_{-\mathbf{k}} \hat{h}_{i j}|\Omega\rangle.
\end{align}
We know from isotropy that this three-point correlator only depends on the modulus of the momentum. Thus the integral is proportional to $M^{ij}\delta_{ij}$, which is zero because $M$ is traceless. 

As a result, $\Delta\varepsilon_k$ doesn't contribute to our final result. For exactly the same reason, we can also neglect $V$ defined in the main text.
\section*{Acknowledgement}
We thank Jason Kristiano and
Keisuke Inomata for useful discussions. 
This work is supported in part by the National Key Research and Development Program of China No. 2020YFC2201501, in part by the National Natural Science Foundation of China No. 12475067 and No. 12235019.


\bibliographystyle{utphys}

\bibliography{SYM}

\end{document}